# Tetra-hyperbolic and tri-hyperbolic optical phases in anisotropic metamaterials without magnetoelectric coupling due to hybridization of plasmonic and magnetic Bloch high-k polaritons.

Maxim Durach

Department of Physics and Astronomy, Georgia Southern University, Statesboro, GA 30460, USA

mdurach@georgiasouthern.edu

**Abstract.** In this paper we reveal the physics behind the formation of tri- and tetra-hyperbolic phases in anisotropic metamaterials without magnetoelectric coupling and describe the anti-crossing splitting phenomenon in the hyperbolic dispersion which arises due to the hybridization of the plasmonic and magnetic Bloch high-k polaritons. This considerably deepens the understanding of the high-k polaritons and the topology of the optical iso-frequency surfaces in k-space and will find applications in optical nano-resolution imaging and emission rate and directivity control. To accomplish this, we develop a range of new techniques of theoretical optics for bianisotropic materials, including the quadratic index of refraction operator method, suitable to study the high-k polaritons with finite indices of refraction and the explicit expression for the characteristic matrix in generic bianisotropic media. We introduce the spatial stratification approach for the electric and magnetic responses of anisotropic homogeneous media to analyze the underlying Bloch waves. We believe that the formalisms developed here can be useful for the researchers in the field of theoretical optics of anisotropic and bianisotropic media in the future.

In the end of this decade the variety in the field of hyperbolic metamaterials and iso-frequency topology in k-space has been drastically expanded and this research direction attracts renewed interest [1-3]. Conventionally, materials in the non-hyperbolic phase have closed iso-frequency k-surfaces of ellipsoidal nature. The plasmonic hyperbolic metamaterials with the extremely anisotropic dielectric permittivity tensor feature open iso-frequency k-surfaces [4] and find numerous applications [5-7]. In 2016 *Kruk et al.* [8] experimentally observed magnetic hyperbolic metamaterials, with hyperbolicity due to the extreme anisotropy in the magnetic permeability tensor. In 2017 *Tuz, Fedorin and Fesenko* [9] predicted that simultaneous extreme anisotropy in dielectric permittivity and magnetic permeability tensors can be observed in non-reciprocal gyrooptic magnetic-semiconductor superlattice resulting in bi-hyperbolic optical phase.

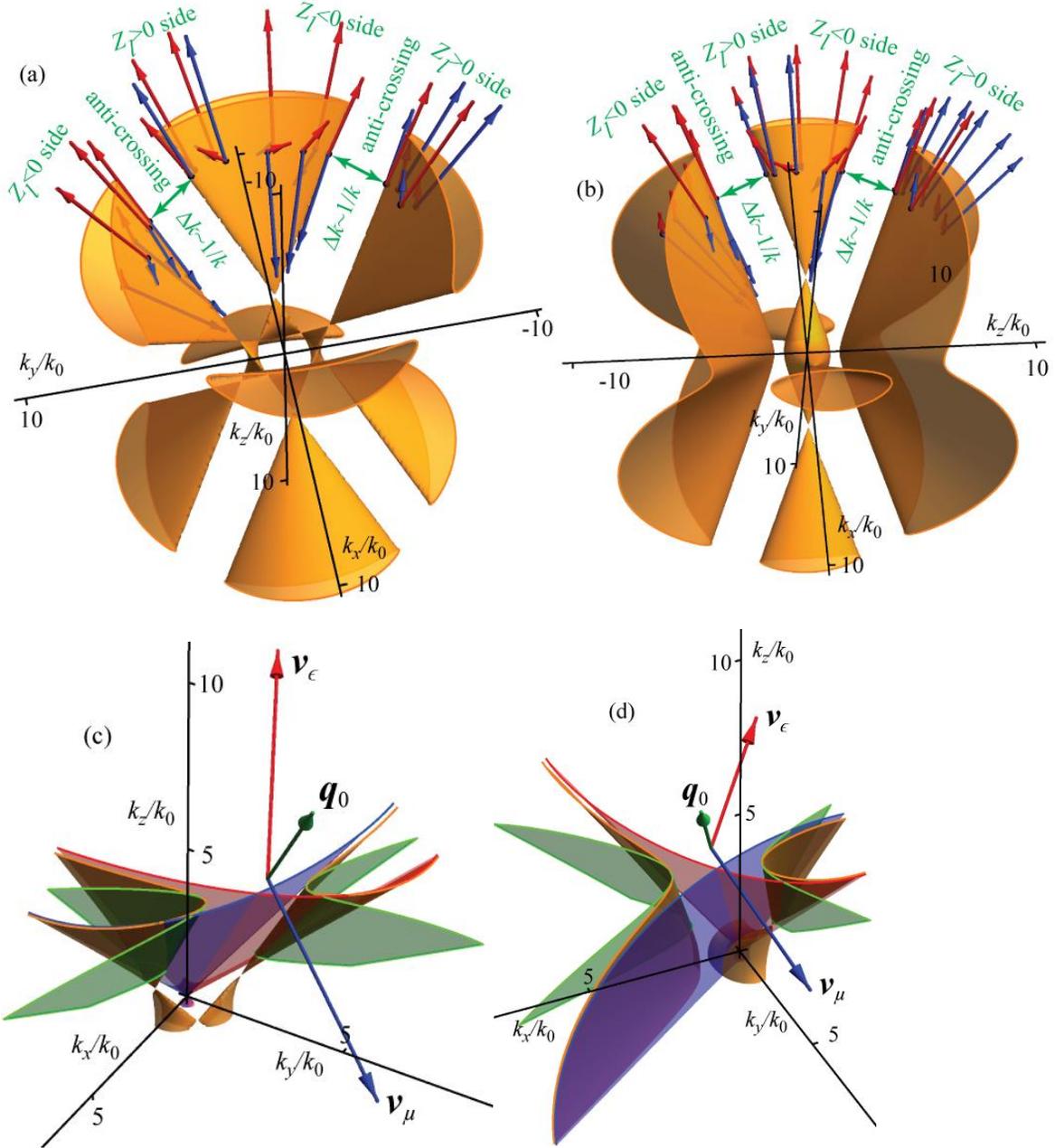

Fig. 1. Iso-frequency k-surfaces for (a) the tetra-hyperbolic-like dispersion in a material with $\hat{\epsilon} = \text{diag}\{-2,-1,1\}$ and $\hat{\mu} = \text{diag}\{1,2,-1\}$ and (b) the tri-hyperbolic-like dispersion in a material with $\hat{\epsilon} = \text{diag}\{-2,-1,1\}$ and $\hat{\mu} = \text{diag}\{-1,2,-1\}$. Some of the anti-crossings are marked in green; the longitudinal impedances $Z_l$ at each side of the anti-crossings are indicated. The electric and magnetic fields of selected plane waves are shown with red and blue vectors respectively. As one can see the fields are in-phase and out-of-phase on different sides of the anti-crossings corresponding to longitudinal impedances of opposite signs. In (c)-(d) the anti-crossings from the k-surfaces (a)-(b) (shown in orange) are compared with the approximation of Eq. (15) (green). The magnetic and electric branches $\tilde{\epsilon}_{33} = 0$ and $\tilde{\mu}_{33} = 0$ are shown as red and blue surfaces with normal vectors $\boldsymbol{v}_\epsilon$ and $\boldsymbol{v}_\mu$ colored correspondingly. The degenerate directions are indicated with the green vectors $\boldsymbol{q}_0$.

At the beginning of 2020 *Durach et al.* [10] introduced the complete taxonomy of bianisotropic optical materials, which include the following topological classes: non-, mono-, bi-, tri-, and tetra-hyperbolic materials. The prefix in the name of each topological class indicates the number of the double cones that the iso-frequency k-surface has in the $k \to \infty$ limit. Technically, the tri- and tetra-hyperbolic phases require magnetoelectric coupling and in the "mono"-anisotropic regime (i. e. without magnetoelectric coupling) only non-, mono-, and bi-hyperbolic phases can be achieved [10]. However, as *Tuz and Fesenko* demonstrate now [11], even the "mono"-anisotropic materials can support iso-frequency topologies, resembling tetra-hyperbolic (see Fig. 3(d) in Ref. [11]), at least for the high-k polaritons with finite k-vectors.

In this paper we explain the phenomenon of formation of tri- and tetra-hyperbolic-like topological phases in media without magnetoelectric coupling by hybridization of plasmonic and magnetic Bloch high-k polaritons which leads to the separation of the 2 double cones of the bi-hyperbolic dispersion into 3 or 4 double cones via anti-crossing whose width is proportional to $\Delta k \propto k^{-1}$ and disappears as $k \to \infty$. The examples of such anti-crossings are marked with green arrows in Figs. 1 (a)-(b) for anisotropic materials with $\hat{\epsilon} = \text{diag}\{-2, -1, 1\}$ and $\hat{\mu} = \text{diag}\{1, 2, -1\}$ in panel (a) and $\hat{\epsilon} = \text{diag}\{-2, -1, 1\}$ and $\hat{\mu} = \text{diag}\{-1, 2, -1\}$ in panel (b).

Such separation of the dispersion cones for the high-k polaritons with finite $k \gg k_0$ is extremely relevant in practice, even though it disappears for $k \to \infty$, since the hyperbolic regimes are realized not in homogeneous materials, but in metamaterials with engineered anisotropic or bianisotropic optical properties achieved by designing subwavelength meta-atom unit cells with finite dimensions $\sim d$. Correspondingly, the highest electromagnetic k-vectors that can be achieved are on the order of $k_{\text{max}} \sim \frac{\pi}{d}$ [12-14]. Therefore, the hybridization of the high-k polaritons gives an opportunity to access the tetra- and tri-hyperbolic regimes for finite $k_{\text{max}} \sim \frac{\pi}{d}$ in anisotropic metamaterials even without magnetoelectric coupling.

To analyze the parameters of the anti-crossing we modify the *index of refraction operator* method for the tangential fields [10] into the *quadratic index of refraction operator* method for the longitudinal fields, which is especially suitable to study the high-k polaritons with finite $n \gg 1$. We further develop our index of refraction operator method and obtain the explicit expression for the *characteristic matrix in bianisotropic materials*. To reveal the physical nature of the anti-crossing splitting of the iso-frequency k-surface cones as the hybridization of the plasmonic and magnetic Bloch high-k polaritons we introduce the spatial stratification approach to spatially separate the dielectric and the magnetic responses inside the anisotropic homogeneous effective medium to get access to the physics of the underlying Bloch waves and their iso-frequency k-surfaces.

The homogeneous media featuring bianisotropic optical response are described by the 6x6 material parameters matrix $\hat{M}$, such that

$$\begin{pmatrix} \boldsymbol{D} \\ \boldsymbol{B} \end{pmatrix} = \hat{M} \begin{pmatrix} \boldsymbol{E} \\ \boldsymbol{H} \end{pmatrix} = \begin{bmatrix} \hat{\epsilon} & \hat{X} \\ \hat{Y} & \hat{\mu} \end{bmatrix} \begin{pmatrix} \boldsymbol{E} \\ \boldsymbol{H} \end{pmatrix} \quad (1)$$

We start by considering Maxwell's equations in the homogeneous medium for the fields $\hat{\Gamma} = \begin{pmatrix} \boldsymbol{E} \\ \boldsymbol{H} \end{pmatrix}$ in plane waves $\hat{\Gamma}(\boldsymbol{r},t) = \hat{\Gamma} e^{i(k_x x + k_y y + k_z z - \omega t)}$, with k-vectors $\boldsymbol{k} = (k_x, k_y, k_z)$

$$\hat{Q}\hat{\Gamma} = -ik_0 \hat{M}\hat{\Gamma}, \quad \hat{Q} = \begin{bmatrix} \hat{0} & \hat{R} \\ -\hat{R} & \hat{0} \end{bmatrix}, \quad \hat{R} = \begin{pmatrix} 0 & -ik_z & ik_y \\ ik_z & 0 & -ik_x \\ -ik_y & ik_x & 0 \end{pmatrix} \quad (2)$$

As was described in [10], the closed-form solution of this equation can be obtained by transforming Eq. (2) into the coordinate system where the k-vector is along one of the axes (z-axis in our notations): $\widetilde{M} = \hat{T}_6 \hat{M} \hat{T}_6^{-1}$, where $\hat{T}_6$ is the corresponding rotation matrix. Breaking down the rotated matrix $\widetilde{M}$ into the following matrixes

$$\widetilde{M}_\| = \begin{pmatrix} \tilde{\epsilon}_{11} & \tilde{\epsilon}_{12} & \tilde{X}_{11} & \tilde{X}_{12} \\ \tilde{\epsilon}_{21} & \tilde{\epsilon}_{22} & \tilde{X}_{21} & \tilde{X}_{22} \\ \tilde{Y}_{11} & \tilde{Y}_{12} & \tilde{\mu}_{11} & \tilde{\mu}_{12} \\ \tilde{Y}_{21} & \tilde{Y}_{22} & \tilde{\mu}_{21} & \tilde{\mu}_{22} \end{pmatrix}, \widetilde{M}_{\|,z} = \begin{pmatrix} \tilde{\epsilon}_{13} & \tilde{X}_{13} \\ \tilde{\epsilon}_{23} & \tilde{X}_{23} \\ \tilde{Y}_{13} & \tilde{\mu}_{13} \\ \tilde{Y}_{23} & \tilde{\mu}_{23} \end{pmatrix}, \hat{P} = \hat{P}^{-1} = \begin{pmatrix} 0 & 0 & 0 & -1 \\ 0 & 0 & 1 & 0 \\ 0 & 1 & 0 & 0 \\ -1 & 0 & 0 & 0 \end{pmatrix},$$

$$\widetilde{M}_{z,\|} = \begin{pmatrix} \tilde{\epsilon}_{31} & \tilde{\epsilon}_{32} & \tilde{X}_{31} & \tilde{X}_{32} \\ \tilde{Y}_{31} & \tilde{Y}_{32} & \tilde{\mu}_{31} & \tilde{\mu}_{32} \end{pmatrix}, \widetilde{M}_z = \begin{pmatrix} \tilde{\epsilon}_{33} & \tilde{X}_{33} \\ \tilde{Y}_{33} & \tilde{\mu}_{33} \end{pmatrix}, \quad (3)$$

the Maxwell's equations for the transverse and longitudinal fields $\widetilde{\Gamma}_\| = (\tilde{E}_x, \tilde{E}_y, \tilde{H}_x, \tilde{H}_y)$ and $\widetilde{\Gamma}_z = (\tilde{E}_z, \tilde{H}_z)$ can be rewritten as

$$k\hat{P}\widetilde{\Gamma}_\| = k_0 (\widetilde{M}_{\|,z} \widetilde{\Gamma}_z - \widetilde{M}_\| \widetilde{\Gamma}_\|) \quad (4)$$

$$\widetilde{\Gamma}_z = \widetilde{M}_z^{-1} \widetilde{M}_{z,\|} \widetilde{\Gamma}_\| \quad (5)$$

From this the direction-dependent *index of refraction operator* can be introduced [10] $\hat{N} = \hat{P}^{-1}(\widetilde{M}_{\|,z} \widetilde{M}_z^{-1} \widetilde{M}_{z,\|} - \widetilde{M}_\|)$, such that its 4 eigenvalues are the indexes of refraction $n(\theta, \varphi) = k/k_0$ for a specific direction $(\theta, \varphi)$ and eigenvectors are $\widetilde{\Gamma}_\|$:

$$\widehat{N}\widetilde{\Gamma}_{\parallel} = n\widetilde{\Gamma}_{\parallel}. \tag{6}$$

The eigenproblem Eq. (6) according to the Faddeev–LeVerrier algorithm [15] leads to the quartic dispersion equation [10]

$$n^4 - \mathrm{tr}(\widehat{N})\, n^3 - \xi n^2 - \zeta n + \det(\widehat{N}) = 0, \tag{7}$$

with coefficients $\xi = \frac{1}{2}\left(\mathrm{tr}(\widehat{N}^2) - \mathrm{tr}(\widehat{N})^2\right)$ and $\zeta = \frac{1}{6}\left(2\,\mathrm{tr}(\widehat{N}^3) - 3\,\mathrm{tr}(\widehat{N}^2)\,\mathrm{tr}(\widehat{N}) + \mathrm{tr}(\widehat{N})^3\right)$.

This formalism provides the parameters of the 4 permitted plane waves in an arbitrary direction $(\theta, \varphi)$ in a generic bianisotropic material. Let us re-express Eq. (5) by excluding the transverse fields $\widetilde{\Gamma}_{\parallel}$ using Eq. (4)

$$\widehat{N}_q \widetilde{\Gamma}_z = 0, \quad \widehat{N}_q = n^2 \widetilde{M}_{\parallel,z}^{-1} \widehat{P} \widetilde{M}_{z,\parallel}^{-1} \widetilde{M}_z + n \widetilde{M}_{\parallel,z}^{-1} \widetilde{M}_{\parallel} \widetilde{M}_{z,\parallel}^{-1} \widetilde{M}_z - \widehat{1} \tag{8}$$

Eq. (8) is a reformulation of the linear index of refraction method for transverse fields $\widetilde{\Gamma}_{\parallel}$ with the 4x4 index of refraction operator $\widehat{N}$ of Eq. (6) into the quadratic eigenvalue problem [16] for the longitudinal fields $\widetilde{\Gamma}_z$ with the 2x2 quadratic index of refraction operator $\widehat{N}_q$.

To study the high-k polaritons it is convenient to rewrite Eq. (8) as

$$\widehat{Z}\widetilde{\Gamma}_z = 0, \quad \widehat{Z} = \left(\widetilde{M}_z - \widetilde{M}_{z,\parallel} n^{-1} \left(\widehat{P} + n^{-1}\widetilde{M}_{\parallel}\right)^{-1} \widetilde{M}_{\parallel,z}\right) \tag{9}$$

The condition $\det \widehat{Z} = 0$ is equivalent to the dispersion equation Eq. (7), while the kernel of $\widehat{Z}$ corresponds to the longitudinal fields $\widetilde{\Gamma}_z$. In the limit $n = k/k_0 \to \infty$ the operator $\widehat{Z} \to \widetilde{M}_z$ and the dispersion equation becomes

$$\det \widehat{Z} = \tilde{\epsilon}_{33}\tilde{\mu}_{33} - \tilde{X}_{33}\tilde{Y}_{33} = h(\boldsymbol{q}) = (\boldsymbol{q}^T \hat{\epsilon} \boldsymbol{q})(\boldsymbol{q}^T \hat{\mu} \boldsymbol{q}) - (\boldsymbol{q}^T \widehat{X} \boldsymbol{q})(\boldsymbol{q}^T \widehat{Y} \boldsymbol{q}) = 0, \tag{10}$$

where $h(\boldsymbol{q})$ is the *high-k characteristic function* introduced in [10] as the high-k limit of the iso-frequency surface for plane waves in bianisotropic materials and $\boldsymbol{q} = \widehat{\boldsymbol{k}} = \frac{\boldsymbol{k}}{k}$ is the unit vector in the k-direction.

In absence of the magnetoelectric coupling $\widehat{X} = \widehat{Y} = 0$ the high-k characteristic function $h(\boldsymbol{q}) = (\boldsymbol{q}^T \hat{\epsilon} \boldsymbol{q})(\boldsymbol{q}^T \hat{\mu} \boldsymbol{q})$ corresponds to the product of 2 double cones $(\boldsymbol{q}^T \hat{\epsilon} \boldsymbol{q}) = 0$ and $(\boldsymbol{q}^T \hat{\mu} \boldsymbol{q}) = 0$ if both dielectric permittivity tensor $\hat{\epsilon}$ and $\hat{\mu}$ are extremely anisotropic, i.e. have principle values of different signs. This corresponds to a bi-hyperbolic regime and, at least in the $k \to \infty$ limit, the tri- and tetra-hyperbolic regimes are inaccessible without magnetoelectric coupling. Nevertheless, the $(\boldsymbol{q}^T \hat{\epsilon} \boldsymbol{q}) = 0$ and $(\boldsymbol{q}^T \hat{\mu} \boldsymbol{q}) = 0$ cones may intersect forming the *degenerate directions* in which two kinds of the high-k polaritons with different polarizations can propagate simultaneously. In this paper we show that although these degenerate high-k polaritons do not interact and simply cross in the k-space in the $k \to \infty$ limit, they *do* interact at finite high k values, which results in anti-crossing splitting of the high-k cones and the formation of the tri- and tetra-hyperbolic regimes within the Brillouin zone.

For high but finite $n \gg 1$, the matrix $\hat{Z}$ can be represented as

$$\hat{Z}_{n \gg 1} \approx \widetilde{M}_z - n^{-1} \widetilde{M}_{z,\|} \hat{P}^{-1} \widetilde{M}_{\|,z} \tag{11}$$

and is well suited to study the anti-crossing splitting of the iso-frequency k-surface in the degenerate directions. For clarity, below we consider materials in which both $\hat{\epsilon}$ and $\hat{\mu}$ have the same principle axes and we direct the unrotated coordinate system along those axes. Correspondingly, the dielectric permittivity and magnetic permeability is given by diagonal tensors $\hat{\epsilon} = \text{diag}\{\epsilon_1, \epsilon_2, \epsilon_3\}$ and $\hat{\mu} = \text{diag}\{\mu_1, \mu_2, \mu_3\}$. We assume that this material is bi-hyperbolic and both $\hat{\epsilon}$ and $\hat{\mu}$ have principal values of different signs. In this case $\tilde{\epsilon}_{33} = (\boldsymbol{q}^T \hat{\epsilon} \boldsymbol{q}) = \epsilon_1 q_x^2 + \epsilon_2 q_y^2 + \epsilon_3 q_z^2$, $\tilde{\mu}_{33} = (\boldsymbol{q}^T \hat{\mu} \boldsymbol{q}) = \mu_1 q_x^2 + \mu_2 q_y^2 + \mu_3 q_z^2$. The condition $h(\boldsymbol{k}) = \tilde{\epsilon}_{33} \tilde{\mu}_{33} = 0$ is the equation of two double cones which correspond to the magnetic and electric branches of the bi-hyperbolic dispersion $\tilde{\mu}_{33} = 0$ and $\tilde{\epsilon}_{33} = 0$ respectively [10].

It is convenient to introduce vectors $\boldsymbol{\epsilon} = (\epsilon_1, \epsilon_2, \epsilon_3)$ and $\boldsymbol{\mu} = (\mu_1, \mu_2, \mu_3)$. The two branches of the bi-hyperboloid dispersion $\tilde{\mu}_{33} = 0$ and $\tilde{\epsilon}_{33} = 0$ in the $n \to \infty$ limit correspond to the vector $\boldsymbol{u} = (q_x^2, q_y^2, q_z^2)$ being orthogonal to vectors $\boldsymbol{\epsilon}$ and $\boldsymbol{\mu}$, i.e. $\boldsymbol{u} \cdot \boldsymbol{\epsilon} = 0$ for the electric and $\boldsymbol{u} \cdot \boldsymbol{\mu} = 0$ for the magnetic branch. The electric and magnetic branches overlap along the *degenerate directions*, where both $\tilde{\epsilon}_{33} = 0$ and $\tilde{\mu}_{33} = 0$, or equivalently $\boldsymbol{u}$ is orthogonal to both $\boldsymbol{\epsilon}$ and $\boldsymbol{\mu}$. This is possible if $\boldsymbol{u}$ can be represented as $\boldsymbol{u} = \alpha[\boldsymbol{\epsilon} \times \boldsymbol{\mu}]$, where $\alpha$ is for normalization to maintain $q_x^2 + q_y^2 + q_z^2 = 1$, $\alpha = (\sum_i [\boldsymbol{\epsilon} \times \boldsymbol{\mu}]_i)^{-1}$. Since all the components of $\boldsymbol{u}$ are required to be positive, the degenerate directions are present in a material only if all the components of the vector $[\boldsymbol{\epsilon} \times \boldsymbol{\mu}]$ have the same signs and correspond to unit vectors $\boldsymbol{q}_0 = (q_{0z}, q_{0z}, q_{0z})$ where $q_{0i} = \pm\sqrt{(\alpha[\boldsymbol{\epsilon} \times \boldsymbol{\mu}]_i)}$.

The dispersion of the high-k modes can be obtained from investigation of the matrix $\hat{Z}_{n \gg 1}$ from Eq. (11), which turns into

$$\hat{Z}_{n \gg 1} = \begin{pmatrix} \tilde{\epsilon}_{33} & An^{-1} q_x q_y q_z \\ An^{-1} q_x q_y q_z & \tilde{\mu}_{33} \end{pmatrix}, \tag{12}$$

where $A = \epsilon_1(\mu_2 - \mu_3) + \epsilon_2(\mu_3 - \mu_1) + \epsilon_3(\mu_1 - \mu_2)$. The high-k dispersion of Eq. (10) can be extended to the finite $n \gg 1$ as

$$\det \hat{Z}_{n \gg 1} = h(\boldsymbol{q}) - A^2 n^{-2} q_x^2 q_y^2 q_z^2 = 0 \tag{13}$$

Near the degenerate directions $\boldsymbol{q}_0$ we introduce the normals to the electric branch $\tilde{\epsilon}_{33} = 0$ - $\boldsymbol{v}_\epsilon = (q_{x0}\epsilon_1, q_{y0}\epsilon_2, q_{z0}\epsilon_3)$, and the magnetic branch $\tilde{\mu}_{33} = 0$ - $\boldsymbol{v}_\mu = (q_{x0}\mu_1, q_{y0}\mu_2, q_{z0}\mu_3)$. Then we Taylor-expand the high-k characteristic in the vicinity of the degenerate directions as

$$h(\boldsymbol{q}_0 + \boldsymbol{\delta q}) = 4(\boldsymbol{v}_\epsilon \boldsymbol{\delta q})(\boldsymbol{v}_\mu \boldsymbol{\delta q}) \tag{14}$$

From this we obtain the equation for the anti-crossing splitting of the high-k waves around the degenerate directions

$$4(\boldsymbol{v}_\epsilon \boldsymbol{\delta q})(\boldsymbol{v}_\mu \boldsymbol{\delta q}) = A^2 n^{-2} q_{x0}^2 q_{y0}^2 q_{z0}^2 \tag{15}$$

which is the equation of a hyperbola in curvilinear coordinates with basis $(\boldsymbol{v}_\epsilon, \boldsymbol{v}_\mu)$, where $\boldsymbol{q}_0 \boldsymbol{v}_\epsilon = \boldsymbol{q}_0 \boldsymbol{v}_\mu = 0$, but in general $\boldsymbol{v}_\epsilon \boldsymbol{v}_\mu \neq 0$. The major axis of this hyperbola is proportional to $n^{-1}$ and is characterizing the splitting between the high-k cones which disappears for $k \to \infty$ with the hyperbolic anti-crossing turning into crossing of the magnetic and electric branches of the $h(\boldsymbol{q}) = 0$ dispersion [Eq. (10)]. This is illustrated in the Fig. 1. In the panels Fig. 1 (a)-(b) we show the tetra-hyperbolic-like dispersion for $\hat{\epsilon} = \text{diag}\{-2, -1, 1\}$ and $\hat{\mu} = \text{diag}\{1, 2, -1\}$ and tri-hyperbolic-like dispersion for $\hat{\epsilon} = \text{diag}\{-2, -1, 1\}$ and $\hat{\mu} = \text{diag}\{-1, 2, -1\}$ and indicate two of the anti-crossing splittings in each panel. In Figs. 1(c)-(d) we demonstrate the effectiveness of the approximation in Eq. (15) at the anti-crossings.

The splitting is related to the lifting of the polarization degeneracy present in the $k \to \infty$ limit. In this limit the polarization of the purely longitudinal fields of the magnetic and electric branches of the dispersion is characterized by the different longitudinal impedances $Z_l = \frac{E_z}{H_z}$. It is equal to zero $Z_l = 0$ for the magnetic branch and is divergent $Z_l^{-1} = 0$ for the electric branch [10]. If the high-k waves with finite refraction indices $n \gg 1$ are considered, the longitudinal impedance at the anti-crossing is according to Eqs. (9)-(12)

$$Z_l = -\frac{An^{-1} q_{x0} q_{y0} q_{z0}}{\tilde{\epsilon}_{33}} = -\frac{\tilde{\mu}_{33}}{An^{-1} q_{x0} q_{y0} q_{z0}} \tag{16}$$

While the quantity $An^{-1} q_{x0} q_{y0} q_{z0}$ is constant at a given anti-crossing, $\tilde{\epsilon}_{33}$ and $\tilde{\mu}_{33}$ change sign when passing through $\tilde{\epsilon}_{33} = 0$ and $\tilde{\mu}_{33} = 0$ cones, which means that the longitudinal impedances $Z_l = \frac{E_z}{H_z}$ at different sides of the anti-crossings are of opposite signs as indicated at the anti-crossings in Figs 1(a)-(b). This corresponds to in-phase and out-of-phase electric and magnetic fields on the different sides of the anti-crossings which are also shown in Figs 1(a)-(b).

To get an insight into the nature of the anti-crossing in the k-space we consider a periodic structure with period $d$ in which the magnetic and electric responses of the homogeneous materials discussed above are stratified along the z-axis as is shown in Fig. 2(a). The dielectric and magnetic layers have equal volumetric fractions. The electric permittivities and magnetic permeabilities in the period are

$$\hat{\epsilon}_e = \mathrm{diag}\{2\epsilon_1 - 1, 2\epsilon_2 - 1, (2\epsilon_3^{-1} - 1)^{-1}\} = \mathrm{diag}\{\epsilon_{1e}, \epsilon_{2e}, \epsilon_{3e}\}, \hat{\mu}_e = \hat{1} \quad (17)$$
$$\hat{\epsilon}_m = \hat{1}, \hat{\mu}_m = \mathrm{diag}\{2\mu_1 - 1, 2\mu_2 - 1, (2\mu_3^{-1} - 1)^{-1}\} = \mathrm{diag}\{\mu_{1m}, \mu_{2m}, \mu_{3m}\} \quad (18)$$

for the dielectric layer and magnetic layers respectively, such that the effective medium describing the periodic structure corresponds to the homogeneous materials described above

$$\widehat{M}_\| = \frac{1}{2}\big(\widehat{M}_\|^e + \widehat{M}_\|^m\big), \widehat{M}_z^{-1} = \frac{1}{2}\big((\widehat{M}_z^e)^{-1} + (\widehat{M}_z^m)^{-1}\big)$$

Both the dielectric and magnetic layers are assumed mono-hyperbolic.

In Fig. 2(b) we show the dispersion in the individual dielectric and magnetic layers (brown and green) in comparison with the homogeneous effective medium dispersion (orange) for $k < 10k_0$. The exact dispersion for the Bloch waves in the periodic structure is also shown in Fig. 2(b) as red dots calculated for $k_0 d = 0.1$ and one can see the very close correspondence between the Bloch wave dispersion and the effective medium approximation in the $k < 10k_0$ range. However, for larger k-vectors the Bloch wave dispersion and the effective medium approximation deviate considerably as can be seen from Fig. 2(c), where the dispersion is plotted up to the Brillouin-zone edge $n = \pi/(k_0 d)$.

To calculate the 1D Bloch iso-frequency k-surfaces shown in Figs. 2(b)-(d) we found the characteristic matrixes $\widehat{\Delta}_e$ and $\widehat{\Delta}_m$ of the individual layers. To accomplish this, we modify Eqs. (2)-(6) by transforming $ik_z \to \partial_z$ and finding the z-derivative of the transverse to the z-axis fields

$$\frac{\partial}{\partial z}\Gamma_\| = ik_0 \widehat{\Delta}\Gamma_\| \quad (18)$$

$$\Gamma_z = \widehat{M}_z^{-1} \cdot (\widetilde{M}_{z,\|} + \hat{q})\Gamma_\| \quad (19)$$

As a result, we find an explicit expression for the *characteristic matrix in a generic bianisotropic material* $\widehat{\Delta}$:

$$\widehat{\Delta} = \hat{P}^{-1} \cdot \Big(\big(\widehat{M}_{\|,z} + \hat{q}^T\big) \cdot \widehat{M}_z^{-1} \cdot \big(\widehat{M}_{z,\|} + \hat{q}\big) - \widehat{M}_\|\Big), \quad (20)$$

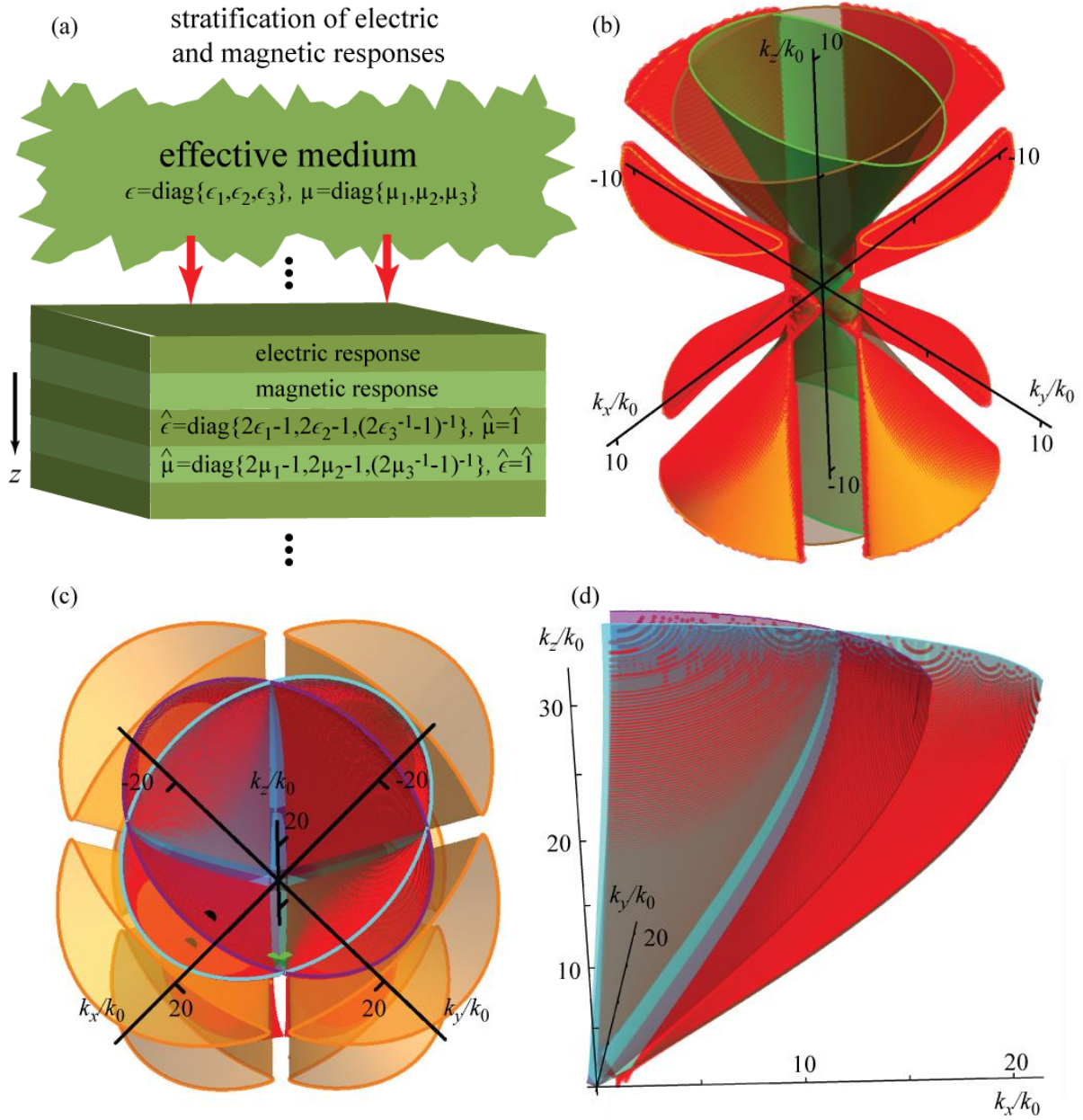

Fig. 2. (a) Illustration of the stratification approach in which the electric and magnetic responses of a homogeneous medium are stratified to get an insight into the Bloch wave behavior underlying the optical response of the original homogeneous medium. (b) The iso-frequency k-surfaces for $k < 10k_0$ of the homogeneous medium (orange) with $\hat{\epsilon} = \text{diag}\{-2,-1,1\}$ and $\hat{\mu} = \text{diag}\{1,2,-1\}$ [same as in Figs. 1(a), (c)]; the electric layers (brown) with $\hat{\epsilon} = \text{diag}\{-5,-3,1\}, \hat{\mu} = \hat{1}$; the magnetic layers (green) with $\hat{\epsilon} = \hat{1}, \hat{\mu} = \text{diag}\{1,3,-1/3\}$ and the Bloch waves dispersion [using Eqs. (22) and (20)] shown with red dots for $k_0 d = 0.1$. (c) The iso-frequency k-surfaces for the same homogeneous medium as in panel (b) and the exact Bloch waves dispersion [using Eqs. (22) and (20)] inside the Brillouin zone. The dispersion of the plasmonic and magnetic Bloch high-k polaritons [Eqs (27)-(28)] is shown in purple and cyan respectively. The black and green dots are in reference to the fields in Fig. 3 below. (d) The exact Bloch waves dispersion [using Eqs. (22) and (20)] in comparison with the dispersion of the plasmonic and magnetic Bloch high-k polaritons [Eqs (27)-(28)] shown in purple and cyan respectively.

There are two differences between the matrix $\widehat{\Delta}$ and the index of refraction operator $\widehat{N}$. First, the matrixes $\widehat{M}_\parallel, \widehat{M}_z, \widehat{M}_{\parallel,z}, \widehat{M}_{z,\parallel}$ are not rotated, i.e. $\widehat{T}_6 = \widehat{1}$ in Eqs. (18)-(20). Second, matrices $\hat{q}$ and $\hat{q}^T$ are introduced into Eqs. (18)-(20) according to

$$\hat{q} = \begin{pmatrix} 0 & 0 & -q_y & q_x \\ q_y & -q_x & 0 & 0 \end{pmatrix}, \qquad (21)$$

The eigenvalues $q_z = q_z(q_x, q_y)$ of the characteristic matrix $\widehat{\Delta}$ provide the shape of the iso-frequency surface identically to the dispersion Eq. (7) or $\det \widehat{Z} = 0$ (see Eqs. (8)-(9)).

Using Eq. (20) as the characteristic matrices of the dielectric and magnetic layers, $\widehat{\Delta}_e$ and $\widehat{\Delta}_m$, we calculate the eigenvalues $q_{Bz}$ of the Bloch transfer matrix $\widehat{\Delta}_B$:

$$\widehat{\Delta}_B = \frac{1}{ik_0 d} \ln \left\{ \exp\left(\frac{ik_0 d}{2} \widehat{\Delta}_e\right) \exp\left(\frac{ik_0 d}{2} \widehat{\Delta}_m\right) \right\}, \qquad (22)$$

which represents the dispersion $q_{Bz} = q_{Bz}(q_x, q_y)$ of the Bloch waves in the stratified structure.

In this paper we are predominantly interested in the high-k waves near the edge of the Brillouin zone and this is where the effective medium approximation considerably diverges from the Bloch waves dispersion found using Eq. (22) [see Fig. 2(c)]. Despite this the effective medium correctly approximates the high-k waves up to $k \sim 10 k_0$, including the anti-crossing regions described by Eqs. (15)-(16). For $k > 10 k_0$ the deviation of the Bloch waves dispersion from the effective medium approximation leads to higher k-vectors for given transverse k-vector components $\boldsymbol{q}_\parallel = (q_x, q_y)$ and narrower anti-crossing gaps as seen in Fig. 3(c), but the tetra-hyperbolic nature of the dispersion is preserved for the Bloch waves.

The key to understanding the tri- and tetra-hyperbolic topology of the iso-frequency surfaces for the Bloch waves is the polarization structure of the Bloch high-k polaritons. In each of the dielectric and magnetic layers [Eq. (17)-(18)] for a given $\boldsymbol{q}_\parallel = (q_x \gg 1, q_y \gg 1)$ there are 4 quasistatic $k \gg k_0$ longitudinally polarized solutions of Eq. (7) [10]:

homogeneous waves $\boldsymbol{E} = \boldsymbol{q} = (\boldsymbol{q}_\parallel, \pm q_z^e)$, with $|\boldsymbol{E}| \gg |\boldsymbol{H}|$ (23a)

inhomogeneous waves $\boldsymbol{H} = \boldsymbol{q} = (\boldsymbol{q}_\parallel, \pm i\kappa_z)$, with $|\boldsymbol{H}| \gg |\boldsymbol{E}|$ (23b)

in the electric layers and

homogeneous waves $\boldsymbol{H} = \boldsymbol{q} = (\boldsymbol{q}_\parallel, \pm q_z^m)$, with $|\boldsymbol{H}| \gg |\boldsymbol{E}|$ (24a)

inhomogeneous waves $\boldsymbol{E} = \boldsymbol{q} = (\boldsymbol{q}_\parallel, \pm i\kappa_z)$, with $|\boldsymbol{E}| \gg |\boldsymbol{H}|$ (24b)

in the magnetic layers with

$$q_z^e = \sqrt{-\frac{\epsilon_{1e}q_x^2 + \epsilon_{2e}q_y^2}{\epsilon_{3e}}}, \quad q_z^m = \sqrt{-\frac{\mu_{1m}q_x^2 + \mu_{2m}q_y^2}{\mu_{3m}}}, \quad \kappa_z = \sqrt{q_x^2 + q_y^2} \tag{25}$$

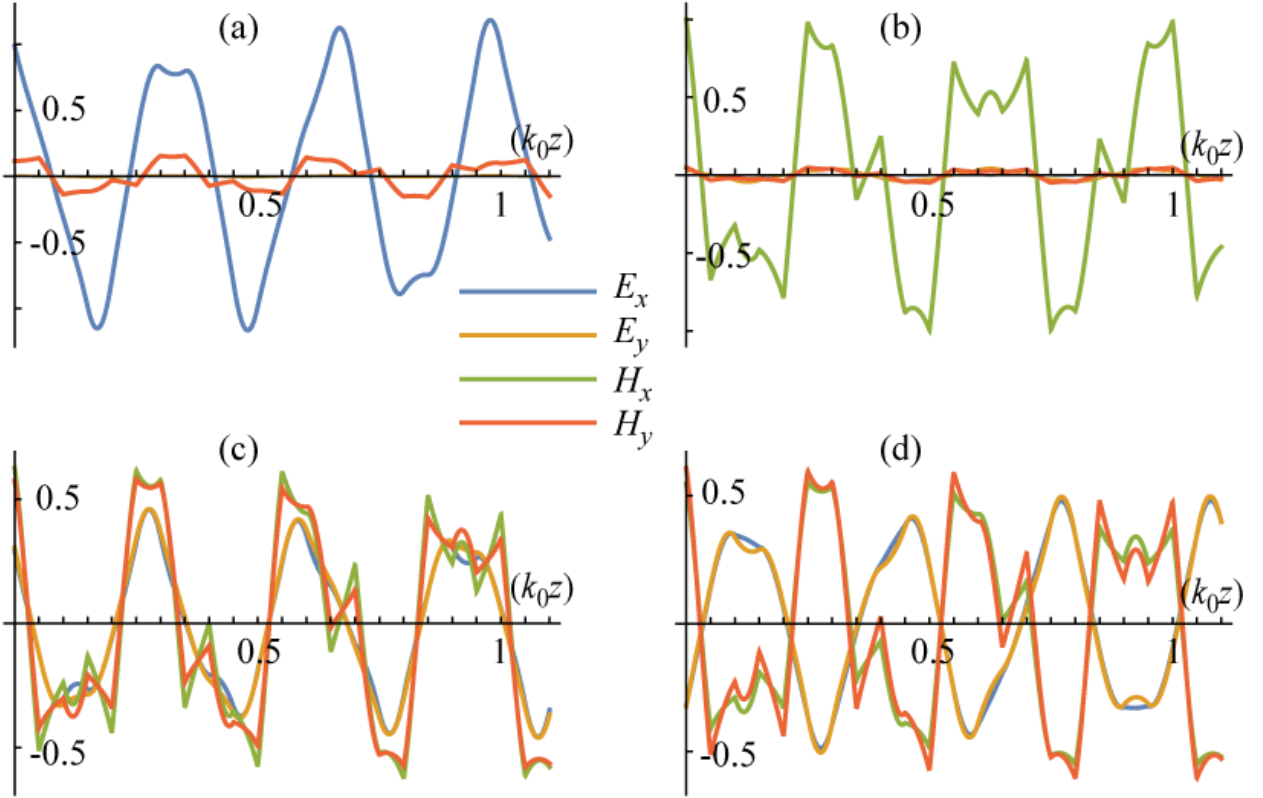

Fig. 3. The fields in the Bloch waves indicated in Fig. 2(c) by the black [panels 3(a)-(b)] and green dots [3(c)-(d)]. One can see the Bloch waves composed of (a) predominantly electric fields $\Gamma_{\parallel e}$ of the plasmonic Bloch high-k polaritons, (b) magnetic fields $\Gamma_{\parallel m}$ of the magnetic Bloch high-k polaritons, and (c)-(d) their linear combinations $\Gamma_{\parallel m} \pm \Gamma_{\parallel e}$. The field components are color-coded as is indicated in the figure.

These waves could be combined into two branches of the plasmonic and magnetic Bloch high-k polaritons. For the plasmonic wave $|\boldsymbol{E}| \gg |\boldsymbol{H}|$, and since it is quasistatic, we can use the electric potential $\varphi_e$, such that $\boldsymbol{E} = -ik_0\boldsymbol{q}\varphi_e$ or the vector potential $\boldsymbol{A} = -\boldsymbol{q}\varphi_e$, which satisfies

$$\varphi_e = \begin{cases} Ae^{ik_0 q_z^e z} + Be^{-ik_0 q_z^e z}, \text{in electric layer} \\ Ce^{k_0\kappa_z z} + Be^{-k_0\kappa_z z}, \text{in magnetic layer} \end{cases} \tag{26}$$

Developing the 2x2 characteristic matrix formalism for the vector $(\varphi_e, \epsilon_{3e}\frac{\partial \varphi_e}{\partial z})$ we obtain the Bloch vector of the plasmonic Bloch high-k polaritons

$$\cos(k_0 q_{Bz}^e d) = \cos\left(\frac{k_0 q_z^e d}{2}\right) \cosh\left(\frac{k_0 \kappa_z d}{2}\right) - \frac{\gamma_e - \gamma_e^{-1}}{2} \sin\left(\frac{k_0 q_z^e d}{2}\right) \sinh\left(\frac{k_0 \kappa_z d}{2}\right) \quad (27)$$

where $\gamma_e = \frac{\epsilon_{3e} q_z^e}{\kappa_z}$.

Similarly, we apply the dual vector potential $\boldsymbol{C}$ as introduced in [17] $\boldsymbol{H} = ik_0\boldsymbol{C}$ (for the ideological basis of dual electromagnetism see Ref [18]). The dual vector potential can be expressed as $\boldsymbol{C} = -\boldsymbol{q}\varphi_m$, where function $\varphi_m$ is similar to the electric potential $\varphi_e$ in Eq. (26), except for the substitutes $q_z^e \to i\kappa_z$ in the electric layer and $\kappa_z \to iq_z^m$ in the magnetic layer. Developing the characteristic matrix formalism for the vector $(\varphi_m, \mu_{3m}\frac{\partial \varphi_m}{\partial z})$ we obtain the Bloch vector of the magnetic Bloch high-k polaritons

$$\cos(k_0 q_{Bz}^m d) = \cos\left(\frac{k_0 q_z^m d}{2}\right) \cosh\left(\frac{k_0 \kappa_z d}{2}\right) - \frac{\gamma_m - \gamma_m^{-1}}{2} \sin\left(\frac{k_0 q_z^m d}{2}\right) \sinh\left(\frac{k_0 \kappa_z d}{2}\right) \quad (28)$$

where $\gamma_m = \frac{\mu_{3m} q_z^m}{\kappa_z}$. The electric and magnetic branches of the Bloch high-k polariton dispersion given by Eqs. (27)-(28) are shown in Figs. 2(c)-(d) as purple and cyan respectively and approximate the exact Bloch wave dispersion (red dots) extremely well. Surprisingly, the discrepancy is negligible even for $k \sim k_0$ and only appears for $k < k_0$, which gives a considerable overlap with the effective medium approximation validity region $k < 10k_0$. The only deviation of the approximation given by Eqs (27)-(28) from the exact Bloch wave dispersion is the absence of the anti-crossing splitting. The reason for this is that Eqs. (27)-(28) describe the uncoupled electric and magnetic Bloch high-k polaritons, while the anti-crossings are due to hybridization between these two types of high-k polaritons as can be seen in Fig. 3.

In Fig. 3 we plot the electromagnetic fields of the high-k Bloch waves in the periodic structure for the k-vectors shown in Fig. 2(c) as black dots away from the anti-crossing near the x-axis and as green dots at the anti-crossing. We see that away from the anti-crossing [Fig. 3(a)-(b)] the fields are of the nature described by Eqs. (23)-(28) with the electric fields $\Gamma_{\parallel e}$ dominating at the electric branch [Fig. 3(a), black dot with smaller $|\boldsymbol{q}_\parallel|$ in Fig. 2(c)] and the magnetic fields $\Gamma_{\parallel m}$ dominating at the magnetic branch [Fig. 3(b), black dot with larger $|\boldsymbol{q}_\parallel|$ in Fig. 2(c)]. The fields at the different sides of the anti-crossing [Fig. 3 (c)-(d)] correspond to the hybridization of the plasmonic and magnetic Bloch high-k polaritons $\Gamma_\| = \Gamma_{\|m} \pm \Gamma_{\|e}$.

In conclusion, it is important to discuss an interesting feature of the exact Bloch wave dispersion shown in Figs. 2(c)-(d). It turns out that the anti-crossings of the Bloch high-k polaritons close and become crossings at the edge of the Brillouin zone $k_{Bz} = \frac{\pi}{d}$, at least for materials with $\widehat{X} = \widehat{Y} = \widehat{0}$ and $\widehat{M}_{z,\|} = \widehat{M}_{\|,z} = \widehat{0}$. In this case the characteristic matrixes $\widehat{\Delta}_e$ and $\widehat{\Delta}_m$ are composed of the off-diagonal blocks $\widehat{\Delta}_{12}$ and $\widehat{\Delta}_{21}$, such that Eq. (18) turns into the system $\frac{\partial}{\partial z}\boldsymbol{E} = ik_0\widehat{\Delta}_{12}\boldsymbol{H}$ and $\frac{\partial}{\partial z}\boldsymbol{H} = ik_0\widehat{\Delta}_{21}\boldsymbol{E}$. This leads to the symmetry with respect to the transformation $k_{Bz} \to -k_{Bz}$, which results in the field change $\boldsymbol{E} \to \boldsymbol{E}, \boldsymbol{H} \to -\boldsymbol{H}$. At the Brillouin zone edge boundary $k_{Bz} = \frac{\pi}{d}$ is equivalent to $-k_{Bz}$ and, therefore, the states $\Gamma_\| = \Gamma_{\|m} \pm \Gamma_{\|e}$ at the different sides of anti-crossing become degenerate and the anti-crossing gap is closed.